\documentstyle[psfig]{l-aa}
\input psfig 
\begin{document}

\thesaurus{03.13.2, 11.03.1, 11.09.03, 12.03.3,  13.25.3}

\title{X-ray structures in galaxy cluster cores\thanks{Based on ROSAT archive data}}

\author{
M. Pierre\inst{1}
\and J.-L. Starck\inst{1}} 
 
\offprints{ M. Pierre    (mpierre@cea.fr)}
 
\institute{CEA Saclay DSM/DAPNIA, Service d'Astrophysique, 
F-91191 Gif sur Yvette}

\date{ accepted for publication}
\maketitle

\begin{abstract}

Using a set of ROSAT HRI deep pointings, we investigate the presence of small-scale structures
in the  central regions of  clusters of galaxies. Our sample comprises 23 objects up to z=0.32, 13 of them known to host a cooling flow.  Structures
are detected and characterized using a wavelet analysis, their statistical significance being assessed by a rigorous treatment of  photon noise.
For all clusters, we present multi-resolution filtered images, restoring structures on all scales and allowing the suppression of noise. 
 We then investigate in detail the geometrical properties of the smallest scale structures. 
Contrary to previous claims, we find very few ``filaments" or point-like features at a 3.7$\sigma$ level, except at the very cluster centers.
Complex cores are conspicuous in at least three  massive cooling flows located at $z = 0.22-0.26$.
From our initial data set we have simulated a redshifted sample, and analyzed it in the same way in order to investigate any instrumental/resolution effect on the detectability of structures. On the one hand, the  topology of the core   down to the limiting resolution appears   to be, at least in our redshift range, indistinguishable between low and high z clusters. On the other hand, external parts seem to be more affected in distant clusters, as indicated by the study
of the ``centroid shift" or position angle variation as a function of radius.
 Peculiar central features and strong isophotal twisting are found in some distant massive cooling flow clusters. All this suggests that X-ray cores  which extend  to a region comparable to the cD envelope should be rather isolated from the rest of the cluster  and  are probably undergoing peculiar physical processes - like ISM/ICM connections  - competing with relaxation.  


\end{abstract}

\section {Introduction}
Structures in clusters of galaxies  constitute one of the most explored areas
of today's observational -- and theoretical -- cosmology. As the
largest bound entities known in the universe, clusters are the ideal tools
to investigate the predictions of the various cosmological theories. Indeed, the way clusters are expected to form in classical hierarchical scenarios, by merging or fragmentation, is strongly dependent on the nature and amount of dark matter present in the universe and is also related to the spectrum of the initial fluctuations of density. Moreover, as pointed out by Richstone et al (1992), density fluctuations continue to grow for longer
in a high density universe than in a low $\Omega$ universe.
All of this should thus be in some way reflected in cluster morpholgies as a function of cosmic time.

From the observational point of view, two practical difficulties arise: (i) structure detection and characterization and (ii) dependence of the detected structures (if any) on the cosmological parameters. Simple empirical questions may be  addressed first, in order to shed light on the topic.  Among them, we would cite: 
Do present day clusters result from the merging
of smaller groups or from the collapse of a single primordial fluctuation? At which stage do clusters start being ``relaxed" (if ever) and what is the influence of the environment? How long is the epoch of cluster formation? Can we detect any difference  between the history of  lens-clusters and non lens-clusters? What triggers or inhibits cooling flows ...?  
Practically, three stages of cluster evolution may be empirically distinguished: (i) a merging (or pre-merging) phase showing two or more distinct close galaxy groups  (as far as can be inferred by their radial velocities) coming into contact, which are also well separated in the X-ray band, (ii) a post-merging phase where the overall galaxy distribution is more compact (but may still show clumps)  corresponding in X-ray to an irregular single area  emission somewhat clumpy on smaller scales and finally  (iii) apparently well-relaxed objects showing an elliptical (or spherical) X-ray morphology. However, it must be kept in mind that a well-relaxed object can ``evolve'' toward a merging phase by further accreting infalling groups.	  
Since phase (ii) and (iii) present far fewer obvious observational signatures than (i) we shall investigate mostly  these latter phases in the present paper.\\

In the history of cluster dynamics studies, optical data were first investigated  for sub-structures in the spatial distribution of cluster galaxies, as the relics of dynamically different cluster constituents. When galaxy clumps
are conspicuous (and not simply projection effects) it appeared non-trivial to isolate them in terms of velocity components. In many cases, merging seems to leave only  imprints as  an increase of the global cluster velocity dispersion, suggesting that some degree of relaxation has already been reached  even though the X-ray image clearly presents  two maxima, and related phenomena (e.g. a radio halo) attest the reality of a recent merging  (cf A2256, Briel et al 1991; A1300, Lemonon et al 1997).
Attention has also been drawn to several cases in which the central dominant galaxy shows a significant peculiar velocity with respect to the cluster mean.
This suggests that, in contrast to the normal assumption, the cD is not at rest in the center of the cluster potential. The presence of a large peculiar velocity is strongly correlated with the presence of substructure in massive sysrem (Bird 1994). An obvious explanation is that the
cluster has not yet reached a state of dynamical equilibrium, probably because
of merging events.
However, in any case, the results are not conclusive regarding  evolution. Moreover, as shown by Dutta (1995), statistics based  on galaxy velocities alone are poor discrimant of $\Omega$ values. 

On the contrary, in the X-ray band,  the hot intra-cluster gas should be the ideal tool to  investigate structures, as projection effects are more easily avoided and as the sound crossing time is much shorter for the gas than for the galaxies. X-ray information will provide an instant view of the cluster potential, if a few basic hypotheses are verified (Schindler 1996). The efficiency of such an approach was already suggested by  EINSTEIN images of nearby clusters (Mohr et al 1993), and ROSAT PSPC (resolution of $\sim 25''$, low background) as well as HRI (resolution of $\sim 5''$) observations revealed a plethora of structures on all scales up to $z \sim 0.4$ (where  fluxes are significant), inspiring great hope for the understanding of cluster formation. In fact, things appeared to be much more complicated as  cluster emission
may be contaminated by unrelated sources (e.g. individual galaxies) and the ability to detect structures
strongly depends on the luminosity of the object and of its distance, because of the limited instrumental
resolution and sensitivity. Phenomenologically, it seems also  almost impossible to isolate true evolution/relaxation processes from external environmental phenomena (whose occurence rate may of course be considered as ``cosmological").  So far, evolution trends are not clear - obvious structured nearby clusters
are indeed observed as well as apparently well-relaxed distant ones. A more systematic and accurate study is needed for  more conclusive results concerning evolution. This is the goal of the present paper which deals with high resolution X-ray observations of a sample of 23 clusters spanning the 0.04-0.32 redshift range where structures are analyzed with a statistically rigorous method. \\

 Here, we focus on the  central region of clusters ($r\leq $ 200 kpc)  
where the matter density (and S/N) is the highest and where crucial phenomena are supposed to take place:
cooling flows, Einstein lensing radius, virialisation as a function of redshift, influence of the cluster
dominant galaxy... For this study, we shall exclusively use ROSAT HRI images in order to take advantage
of the maximum presently available spatial resolution in the X-ray band, as well as to have an homogeneous
data set (for this reason we renounce comparing low-z clusters observed with the PSPC and high-z ones
observed with the HRI).

The paper is organized as follows: in the next Section we present the method used for searching for small
scale structures and, in Sec. 3,  the data set selected from the ROSAT archive data base as well as the processing steps. The results are presented in Sec. 4,  and the discussion  and conclusions are given in Sec. 5

We assume $H_{o}$ = 50 km/s/Mpc and $q_{o}$ = 1/2, coordinates are given in the J2000 system, and position angles are measured counter-clockwise from North throughout this paper.

\section{General  analysis}

In the morphological analysis of X-ray images, which are most often photon noise
dominated, the ability to assess  the statistical significance of possible structures is crucial.
This question is  far from  trivial, as generally only a few photons (down to a few tenths) are present
in individual pixels. Therefore, it is more a question  of small photon number statistics  from the  source   (adding a few photons may  dramatically change the shape of the observed contours) than of background noise, which is generally negligible at the level of interest. Recent controversies on the topic (e.g. Abell 2029, see Sec. 4.2) testify to the technical and scientific  challenge of the problem.   
Buote \& Tsai (1995, 1996) have   considered a power-ratio technique which is a method better suited to the study of global cluster properties. For detecting small scale structures, the most frequent adopted method so far was ellipse or 2D King profile fitting (e.g. White et al 1994, Prestwich et al 1995), 
analyzing residuals after subtraction of the model. However, assessing (i) the reality of the residuals as well as (ii) their characteristic  size, requires a precise knowlegde of the photon statistics at the given image position and adequate  mathematical treatment (e.g. Neumann \& B\"ohringer 1997). This method, even when correctly applied, is nevertheless not well adapted for characterizing the size of the fluctuations (in general very few photons indeed are left in the residuals). Moreover, it often  fails at the very center of the cluster  where  too few pixels are  available for fitting  ellipses. A totally different method, the wavelet analysis, allows a direct isolation of structures as a function of scale (e.g. Slezack et al 1994, Grebenev et al 1995). But so far, the statistical significance of the detected fluctuations has not been discussed, and the method has been mainly  applied on PSPC data having better sensitivity, significantly less instrumental noise but about 5 times lower spatial resolution than the HRI.
With the aim of taking  the maximum benefit of the HRI high angular resolution,   it must be kept in mind that the signal is diluted into more pixels. It thus appears necessary to combine an adequate noise modeling with a wavelet analysis in order to extract all - and not more! - information contained in the images.
The method that we have developed is presented in the dedicated accompanying paper (Starck \& Pierre 1997, hereafter Paper I). For our purpose here, we use it in two steps: (1) detection of the significant structures
(at a given $\sigma$ level)  by a wavelet analysis combined with detailed Poisson noise modeling. The results will be analyzed over an area restricted to the cluster cores at the 4 lowest possible scales ($1\times PSF,~ 2\times PSF, 4\times PSF,~ 8\times PSF$). (2) Restoration of the whole cluster image using the previous multi-resolution analysis. This provides an optimal overall view of the cluster, suppressing noise and restoring structures on different scales.

Finally, since we are investigating structure shapes down to the limiting resolution of the instrument, attention has to be drawn on possible aspect solution effects: residuals in the solution may occasionally produce asymetrical features leading to slightly ellipsoidal images which can have  an amplitude up to 30\% (David et al 1997). In most of the pointings studied here however, point sources are conspicuous, and do not reveal significant asymetry at our filtering level. In addition, 3 of the cluster observations consist of 2 or 3 pointings taken at different epochs: they were individually filtered and present a central morphology very similar to that of the complete observation (Sec. 4.2).

\section{Data set and processing}

We have retrieved from the ROSAT archive data base a sample of 23 clusters, observed with the HRI
with   total exposures greater than 15 ksec. Two clusters having slightly less exposure time were also considered, because they are well known objects, whose morphology has been discussed using other methods. In our final selection, we tried to avoid as much as possible obvious multiple clusters (e.g. A2256) and to keep only bright objects in order to allow us to fully exploit the HRI resolution at least at the cluster center. These two criteria appeared to be rather stringent and we were led to consider ``intermediate cases" in order to have at our disposal a sample of reasonable size.
The objects are listed in Table 1 (hereafter: basic sample).   Sixteen of them appeared to be member of the XBACS sample (the X-ray-brightest Abell-type clusters of galaxies obtained from the ROSAT all-sky survey, Ebeling et al 1996)

\begin{table*}
\caption[]{Basic sample - general characteristics  } 
 \begin{tabular} {||l|l|l|l|l|l|l||} \hline \hline
 Identification & RA (2000) & Dec (2000) & HRI exp. (sec) & $z$ & $z_{eq}$ & $Lum_{eq}$.  \\
\hline \hline
  A 496            &   68.404      &  -13.255      & 14493 & 0.038 & 0.082 &  1.2 \\ 
  A 2572           &  349.626      &  +18.690      & 25028 & 0.040 & 0.085 &  1.2 \\ 
  A 3571           &  206.868      &  -32.867      & 19460 & 0.040 & 0.086 &  1.2 \\ 
  A 85             &   10.462      &   -9.303      & 17308 & 0.052 & 0.115 &  1.3 \\ 
 A 644            &  124.353      &   -7.511      & 18669 & 0.070 & 0.163 &  1.4 \\ 
 A 1423           &  179.323      &  +33.610      & 19030 & 0.076 & 0.179 &  1.4 \\ 
 A 2029           &  227.734      &   +5.745      & 17758 & 0.077 & 0.180 &  1.5 \\ 
 A 2597           &  351.332      &  -12.123      & 17997 & 0.085 & 0.205 &  1.5 \\ 
 A 478            &   63.356      &  +10.465      & 22834 & 0.088 & 0.214 &  1.6 \\ 
 A 2142           &  239.584      &  +27.230      & 19785 & 0.090 & 0.219 &  1.6 \\ 
PKS 0745-191     &  116.880      &  -19.296      & 23385 & 0.103 & 0.260 &  1.7 \\ 
 A 1664           &  195.928      &  -24.245      & 22228 & 0.128 & 0.351 &  2.1 \\ 
RX J2318.3+1843  &  349.591      &  +18.733      & 25028 & 0.160 & 0.480 &  2.5 \\ 
 A 2580           &  350.361      &  -23.208      & 17663 & 0.187 & 0.561 &  2.6 \\ 
 A 383            &   42.014      &   -3.528      & 28074 & 0.187 & 0.561 &  2.6 \\ 
 A 773            &  139.475      &  +51.728      & 16663 & 0.197 & 0.591 &  2.6 \\ 
A 115            &   13.960      &  +26.410      & 50719 & 0.197 & 0.591 &  2.6 \\ 
 MS 0735.6+7421   &  115.434      &  +74.243      & 27164 & 0.216 & 0.648 &  2.6 \\ 
 A 2390           &  328.403      &  +17.695      & 27764 & 0.231 & 0.693 &  2.6 \\ 
 A 2645           &  355.320      &   -9.025      & 35273 & 0.251 & 0.753 &  2.6 \\ 
E 1455+223       &  224.313      &  +22.343      & 14886 & 0.258 & 0.773 &  2.6 \\ 
ZW 3146          &  155.915      &   +4.186      & 26045 & 0.291 & 0.872 &  2.7 \\ 
S 506            &   75.273      &  -24.417      & 37756 & 0.320 & 0.960 &  2.7 \\ 
\hline \hline
\end{tabular}\\
columns:\\
RA and Dec: positions of cluster centers in decimal degrees\\
$z_{eq}$: redshift at which the actual cluster angular distance is multiplied by 2.\\
$Lum_{eq}$ multiplicative factor for the cluster luminosity at $z_{eq}$ (see text)\\
\end{table*}

In Table 2,  further data are presented on the basic sample. The cooling flow  rates are from Edge et al (1992), or from more recent determinations if available. The 2-10 keV fluxes are taken from the XBACS (Ebeling et al 1996)  when available or were computed directly from the images using a standard Raymond \& Smith code (EXSAS), local hydrogen column density (Dickey \& Lockman 1990) and a temperature of 5 keV (except for PKS 745 and ZW 3146 for which we used the published values of 9  and 11 keV  respectively (Allen et al 1996a,b). The fluxes were in turn converted into 2-10 keV luminosities.
 
\begin{table*}
\caption[]{Basic sample - data  } 
 \begin{tabular} {||l|l|l|l|l|l|l|l|l|l|l||} \hline \hline
  Identification  & $z$ & CF rate &  ref & XB & Flux 2-10 keV & Lum 2-10  keV & $I_{cent}$ & $I_{50}$ & W & E    \\  
~ & ~& Mo/yr &  (CF) & ~& $10^{-12} ~erg/sec/cm^{2} $ & $10^{44} erg/sec$ & nb. phot. & nb. phot. & ~ & ~ \\ \hline \hline
  A 496           &  0.038 &  136  & = & 1 & 66.5     &  4.2 &  171 & 3473 & 1 & 1 \\ 
 A 2572          &  0.040 &    0  &   & 2 &  1.9     &  0.1 &   47 &  460 & 2 & 1 \\ 
 A 3571          &  0.040 &  150  & = & 1 & 133.     &  9.2 &   35 & 2236 & 1 & 0 \\ 
 A 85            &  0.052 &  252  & = & 1 & 77.3     &  9.1 &  215 & 3053 & 0 & 1 \\ 
 A 644           &  0.070 &  382  & = & 1 & 40.3     &  8.9 &   36 &  811 & 0 & 0 \\ 
 A 1423          &  0.076 &    0  &   & 1 &  6.1     &  1.6 &   48 &  393 & 1 & 1 \\ 
 A 2029          &  0.077 &  300  & c & 1 & 74.6     & 19.5 &  356 & 3393 & 2 & 1 \\ 
 A 2597          &  0.085 &  590  & c & 1 & 30.6     &  9.9 &  279 & 2416 & 2 & 1 \\ 
 A 478           &  0.088 &  900  & a & 1 & 43.3     & 15.0 &  261 & 2119 & 2 & 1 \\ 
 A 2142          &  0.090 &    0  &   & 1 & 87.5     & 31.6 &  126 & 1399 & 0 & 1 \\ 
 PKS 0745-191    &  0.103 & 1000  & e & 0 & 44.8     & 21.3 &  310 & 2061 & 2 & 1 \\ 
 A 1664          &  0.128 &  260  & g & 1 &  8.0     &  5.9 &  121 &  658 & 2 & 1 \\ 
 RX J2318.3+1843 &  0.160 &    0  &   & 2 &  2.4     &  2.9 &   71 &  528 & 2 & 1 \\ 
 A 2580          &  0.187 &    0  &   & 0 &  2.8     &  4.6 &   63 &  254 & 0 & 1 \\ 
 A 383           &  0.187 &    0  &   & 1 &  6.1     &  9.9 &  260 &  780 & 2 & 1 \\ 
 A 773           &  0.197 &    0  &   & 1 &  7.8     & 14.2 &   12 &   66 & 0 & 1 \\ 
 A 115           &  0.197 &    0  &   & 1 & 11.6     & 21.1 &  209 & 1019 & 2 & 1 \\ 
 MS 0735.6+7421  &  0.216 &  ? & d & 0 &  2.8     &  6.2 &  151 &  410 & 2 & 1 \\ 
 A 2390          &  0.231 &  850  & f & 1 & 13.0     & 32.9 &  186 &  517 & 2 & 1 \\ 
 A 2645          &  0.251 &    0  &   & 0 &  1.1     &  3.3 &   20 &   48 & 0 & 1 \\ 
 E 1455+223      &  0.258 & 1500  & b & 0 &  3.6     & 11.5 &  163 &  404 & 2 & 1 \\ 
 ZW 3146         &  0.291 & 1400  & b & 0 & 10.5     & 43.0 &  481 &  982 & 2 & 1 \\ 
 S 506           &  0.320 &    0  &   & 0 &  0.3     &  1.4 &   18 &   24 & 1 & 0 \\ 
\hline \hline
\end{tabular}\\

columns:\\
XB: 1 if member of the XBACS sample (Ebeling et al 1996). 2 indicates two objects (A2572 \& RXJ2318.3+1843) only distant by 3.2' on the sky   and hardly separated by the Rosat All-Sky Survey (cf Sect. 4.2).\\
W: cluster for which a central structure was detected at all scales with the wavelet analysis (1); 2 stands for clusters having a central structure  greater than 10 pixels. NB A 3571 has been included in the sample because it presents a faint central signal at scale 3 (but not at scales 4 and 5; cf Sect. 4.2). \\
E: cluster suitable for ellipse fitting (down to $I_{central}/10)$  .\\
References (mass flow rates):\\
=) Edge et al 1992\\
a) White et al 1994\\
b) Allen et al 1996a\\
c) Allen \& Fabian 1997 (PSPC)\\ 
d) Donahue \& Stocke  1995\\
e) Allen  et al 1996b \\
f) Pierre et al 1996\\ 
g) Allen et al 1995\\
\end{table*}

Since our method is based on individual photon statistics, it is ideally suited for processing HRI images which can be assumed to have no energy dependence and a flat response (at least within the inner 10 arcmin of the detector). Images (512 $\times$ 512 pixels) were prepared  from the photon event tables that were binned using a  1 arcsec pixel. The PSF is thus reasonably oversampled and in the following structure analysis we shall consider  wavelet plane numbers 3, 4, 5, 6 corresponding respectively to scales of 4, 8, 16, 32 arcsec. The first scale is slightly smaller than the PSF but most of the signal coming from unresolved structures will be maximal at this scale. In the restoration procedure we have discarded contributions from wavelet planes number 1 and 2 and included all those from planes 3 to 7. The statistical significance (see paper I) chosen for investigating the cluster cores is either 3 $\sigma$ or 3.7 $\sigma$ and   we have also systematically ignored ``structures'' that were detected with only one significant pixel. It must be also noted that in Poisson statistics - which is the case for all our images - the characterization of the significance of a signal with a ``$\sigma$'' is meaningless. However we keep this notation in this paper: by analogy with Gaussian statistics, it indicates  the probability of a false detection ($10^{-3}$ and $10^{-4}$ for 3 and 3.7 $\sigma$ respectively).  

In the construction of the images, no selection was set on the 15 HRI energy channels. Because we are investigating structures in cluster cores,  expected to be especially cold, it is important to work on the whole energy range, even though this implies higher sky and background contributions. In any case, however, central intensities are high (from 15 to 150 times the mean background level)  and our filtering process is ideally suited to rejecting background fluctuations. The configuration adopted is thus optimal for searching for sub-structures in cluster cores.

Simultaneously, in order to investigate any evolutionary effect, we have simulated a redshifted sample from the basic sample in the following way: We assumed that clusters are now seen at an angular distance twice as large as their actual one, to which  an equivalent redshift can be associated (column $z_{eq}$ in Table 1). From each photon table, we have randomly selected 1/4 of the events and created images again   having 256 $\times$ 256 pixels,  with a  pixel size of 2 (original) arcsec. The redshifted images can thus be processed exactly in the same way as the original ones, keeping in mind that we have artificially degraded the angular resolution accordingly: wavelet plane number 3 will still be indicative of point-like features.  This has also the advantage to allow us to consider  that the redshifted clusters have been observed with the same exposure time as their ``parent" ones and that the influence of the background is strictly equivalent in both cases. The only slight difference  is that the clusters artificially redshifted in this manner have an intrinsic luminosity somewhat higher than the original ones, because of the distinction between luminosity/angular distances. The ratio between the total luminosity of  redshifted and original clusters is given in Table 1 (column: $Lum_{eq}$). 
With this definition, it is also interesting to remark that morphological results at scale $k$ for a redshifted cluster corresponds to scale $k-1$ for the original object (within statistical fluctuations); this can be visually appreciated by comparing Fig. 6 \& 7.

 In what follows, we have divided the basic sample in two sub-samples (at $z$=0.107): the nearby sample (11, objects $< z >$ = 0.0689) and the distant sample (12 objects, $< z >$ = 0.218).
In the redshifted sample, only the 12 nearest clusters ($ z_{eq}\leq 0.4$) will be useful for the comparison with the basic sample; out of these, the counterpart of the distant basic sample contains 9 objects within $0.107 \leq z_{eq} \leq 0.4$ and has $< z_{eq} >$ = 0.209

The probability of detecting significant structures   is strongly dependent on
the number of photons received in the area of interest. This intensity is, in turn, directly related  to the intrinsic cluster luminosity and distance  as well as to the exposure time.  In the following, where we shall attempt to investigate any correlation between the cluster redshift (and/or its central  luminosity) and the   presence of structures  (a  ``hot" question  in the case of cooling cores), this point is crucial. In order to quantify the influence of the S/N, we have indicated in Table 1 the following numbers of photons: those received in the 10$\times$10 arcsec centered on the maximum of the X-ray emission (column: $I_{cent}$)  and those within a box of half-length   50 kpc - at the cluster's redshift - centered at the same place (columns:  $ I_{50}$).

In order to compare the intrinsic properties of the cluster cores and to be
further able to discuss the influence of photon statistics, a few correlations
have been investigated for both the basic and redshifted samples.

Fig.1 presents the distribution of the sample cluster luminosities as a function of distance: bright and faint objects are found at both ends of the redshift range but, as expected, cooling flow (CF) objects are among the brightest ones.
The distribution of $I_{50}$  as a function of  $z$ is presented on Fig. 2. The clear anti-correlation  observed in the basic sample is  a distance effect. At $z$ larger than $\sim$ 0.107, both basic and redshifted samples show a comparable behavior. 
 The distribution of $I_{cent}$ - the most sensitive parameter for small scale structure detection - is plotted on Fig. 3. The horizontal dotted line  (47 photons) indicates a very conservative lower limit  for such a positive detection, i.e. all objects above the line; i.e. if the central intensity is unresolved, only 47 are sufficient to produce a significant signal in the 3rd wavelet scale. This limit was provided by  cluster 2572 (cf Fig. 6. and Table 2) for which a significant central pointlike feature was found.  In this plot again, for the high redshift bin (except one point: Zw 3146), basic and redshifted distributions are similar. Both Fig. 2 \& 3 indicate thus, that in terms of central S/N the basic and redshifted samples present
equivalent detectability.
Finally, we found a  correlation between redshift and mass flow rate in our sample (Fig. 4), which is discussed in Section 5.1.
   
\begin{figure}
\psfig{file=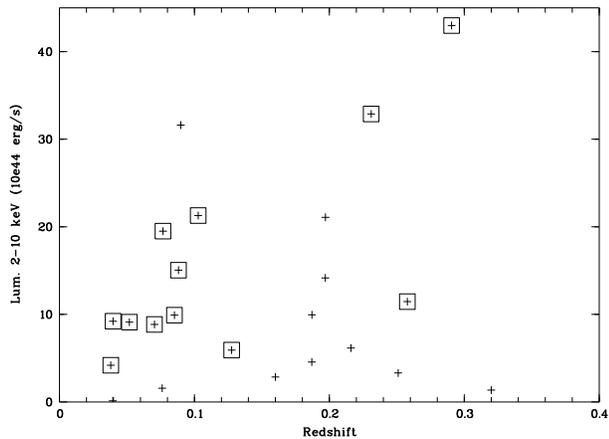,width=9cm,angle=-90}
\caption[] { Total luminosity vs redshift for the sample objects (circles). Squares indicate cooling flow clusters. No correlation is obvious but  cooling objects  tend to be the most luminous ones at any redshift.}
\end{figure}

\begin{figure}
\psfig{file=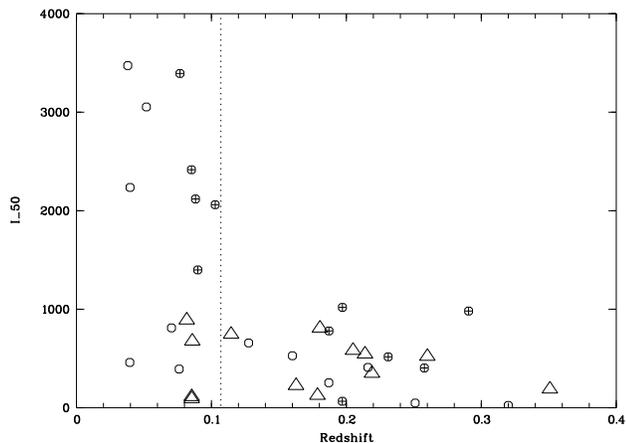,width=9cm,angle=-90}
\caption[] {Number of photons (background subtracted) received within a radius of  50 kpc centered on the maximum of the cluster X-ray emission as a function of $z$. Circles are for the basic sample and crosses indicate objects having $L_{2-10}  \geq 9.3~ 10^{44}~ $ erg/s, which is the median of the sample. Triangles corresponds to the redshifted sample. The vertical dotted line indicates the boundary between low and high $z$ samples. }
\end{figure}

\begin{figure}
\psfig{file=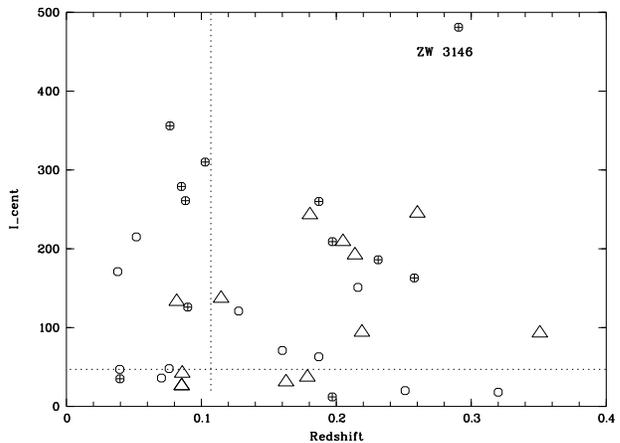,width=9cm,angle=-90}
\caption[] {Number of photons (background subtracted) received in the inner 10$\times$10 arcsec centered on the maximum of the X-ray emission. Same symbols as in Fig. 1. The horizontal dotted line indicates a conservative photon number (47), above which small scale structures - if present - must be detected.}
\end{figure}

\begin{figure}
\psfig{file=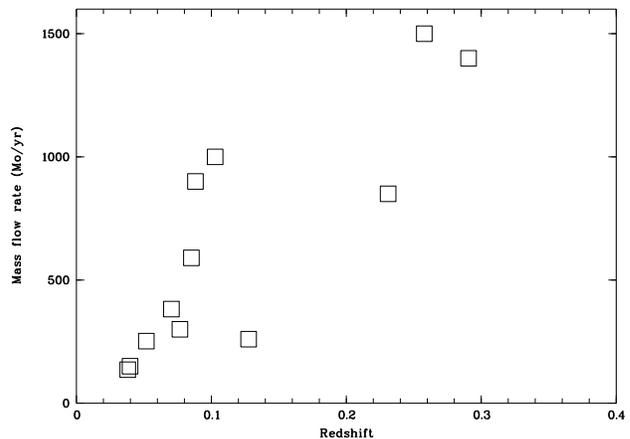,width=9cm,angle=-90}
\caption[] { Mass flow rate vs redshift for the sample clusters identified as cooling objects.}
\end{figure}

\section{Results}

The images restored using the multi-resolution analysis are displayed in Fig. 5
where clusters are ordered according to increasing redshift. For this, only structures above 3.7 $\sigma$ have been taken retained.

\begin{figure}
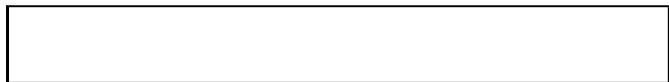

\picplace{1cm}
\caption[] {Cluster images, filtered by our multi-resolution technique. The threshold for structure acceptance was set to 3.7 $\sigma$ (cf Paper I). Contours are in linear scale: first contour is 1/15 of the central value; next ones are equally spaced by 1/7 of the central intensity. The displayed fields are $ 128'' \times 128''$  and axes are in unit of   0.5". Clusters are numbered according to Table 1.}
\end{figure}

Details of the wavelet analysis are presented in Fig. 6. In order to facilitate   visual understanding, the displayed fields all have a size of 200 kpc in the cluster's frame.    Wavelet coefficients  are overlaid with  different line types. For scales of 8" and 16" only the outer contours of the {\em multi-resolution~ support~ image} is drawn, i.e.: the boundary of the image area enclosing all pixels having a significance of at least 3 $\sigma$  for the given scale (see paper I).  For the smallest scale (4") we have also added inner contours, in order to outline possible independent point-like contributions. At this latter scale, the presence of a significant contribution is therefore the sign of a non-resolved emission and, consequently, elongated isophotes can be considered as ``filaments" at our instrumental resolution. 
To be rigorous, in order to unambiguously flag unresolved features, the wavelet coefficient has
to be maximal at scale 3, but this may not practically be true in the case of clusters since a faint small scale
feature may be superimposed on a bright larger one, which would then boost higher order wavelet
coefficients.

Fig. 7 presents similar information for the redshifted sample.

 \begin{figure}
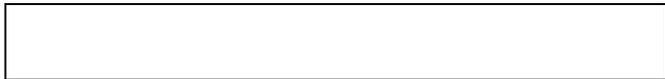

\picplace{1cm}
\caption[] {Wavelet analysis of the sample clusters: contour plots of the wavelet coefficients at the three scales relevant for ROSAT HRI observations. Dashed line: scale 16"; dotted line: scale 8"; solid line: scale 4" ($\sim$ instrumental resolution). For the largest two scales (16", 8"), only the outer contour of the significant area (3$\sigma$) is drawn; for the 4" scale, contours are plotted with an increment
of 0.1 (in wavelet space), the first being also at 3$\sigma$. The displayed fields are 200 kpc $\times$200 kpc (at the cluster's  redshifts) and axes are in unit of   0.5".}
\end{figure}

 \begin{figure}
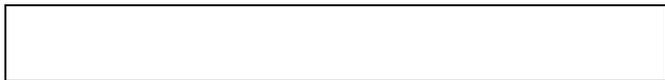

\picplace{1cm}
\caption[] {Same as Fig. 6, for the redshifted sample. The displayed fields are still 200 kpc $\times$ 200 kpc (at the cluster's  redshifts) but axes are in unit of   0.25".}
\end{figure}

\subsection{Overall distributions}
For each   central structure, the following parameters have been derived to characterize the $multi$-$resolution~ support$ at  our 4 relevant scales: centroid position (maximum of the wavelet coefficient), number of significant pixels ($S$), perimeter ($l$), position angle, first order X \& Y moments with respect to the feature main axe and a morphology parameter ($4\pi
S/l^{2}$) quantifying the ``elongation", i.e. departure from a purely circular structure. This latter parameter, hereafter simply ``morphology", characterize the perimeter shape of the significant area of a given scale (i.e. the $support$) and not intensity variations within the area.
This quantitative analysis is restricted to  subsamples containing 17 objects (basic) and 16 objects (redshifted) for which a significant central structure ($3\sigma$) is detected on scales 3, 4, 5, 6; corresponding clusters are flagged in Table 2.\\
 
As shown by Mohr et al (1993), an image centroid which varies with radius indicates the presence of a first harmonic component in the image, which itself implies variations of the center of mass as a function of scale and therefore provides evidence for a significant departure from dynamical equilibrium. This ``center of mass shift" appears also the best means to distinguish between different cosmologies as well as initial power spectra -- at least when considering galaxy optical data (Crone et al 1996).
We have thus investigated a possible centroid shift   within our 3 wavelet scales. Computed shifts are   offsets between centroids  of scale 3-5 and 3-6, the latter for investigating large off-center effects (like infalling groups). 
We found that variations of these quantities are much less significant when considering scales 4-5 and 4-6.
Results are displayed in Fig. 8 \& 9 for the basic and redshifted samples respectively. Wavelets are especially well adapted for this test, as (i) centroid positions on large scales are not affected by the presence of possible pointlike sources and (ii) the positions (maximum of the wavelet coefficient) are filtered values

 \begin{figure}
\psfig{file=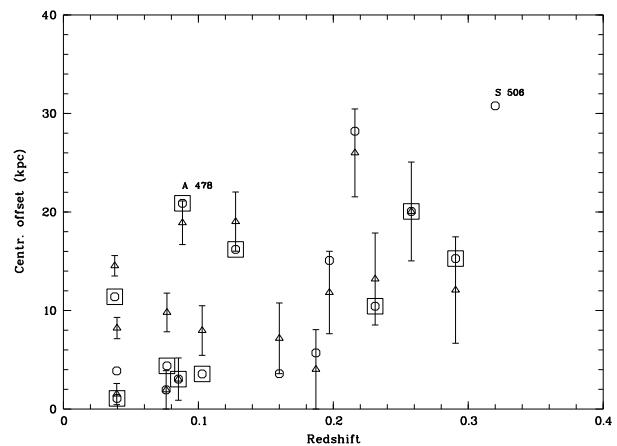,width=9cm,angle=-90}
 \caption[] {Radial variation of the image centroid as a function of redshift for the basic sample. The y axis is the distance between the centroids of scales 3-5 (circles) and scales 3-6 (squares); it is expressed in kpc at the cluster redshift. To each cluster is thus associated a cross and a circle. Error bars correspond to a one pixel (1") uncertainty in distance. Cooling flow clusters are identified with a square. }
\end{figure}

 \begin{figure}
\psfig{file=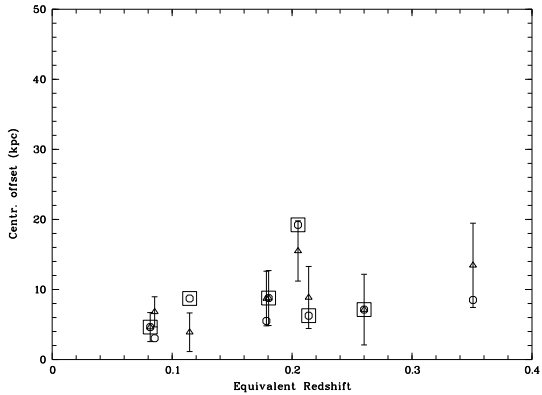,width=8cm,angle=-90}
 \caption[] {Same as Fig. 8, but for the redshifted sample.}
\end{figure}

The comparison between centroid shifts obtained with scales 3-5 and 3-6 show that the estimates are very stable and thus, reliable: the majority of the 2 offset measurements  deviate less than one arcsec. Therefore, although a given scale probes different radial distance as a function of redshift, we are confident that scales 5 \& 6 are well representative of   cluster external regions and  well isolated from the core regions. There are a few exceptions indicating a strong shift between the centroid of scales 5 and 6 and therefore non-relaxed processes in the cluster outskirts as can be seen on the filtered images: in S506 (most distant and very faint object) and in  A478, the offset is due to the fact that the central feature is very elongated, possesses 2 maxima, the brighter one  being clearly off-centered (cf Fig. 6). There is an overall clear trend for an increasing centroid shift with $z$, apparently not connected with the presence of a cooling flow. The corresponding distribution for the redshifted sample shows significantly smaller offsets for all the redshift range.

We have further investigated the shape of cluster cores  and,
for this, considered the properties of the cluster centers at scale 3.
First of all, as shown by Fig. 10, there is no correlation between morphology and central photon number. This confirms the statement  inferred in the previous Section: there are enough photons for most of the clusters to enable a reliable morphology measurement.
Next, Fig. 11  shows the surface of the central feature as a function of $z$.
The complete redshifted sample has been displayed (up to $z$ = 0.8), which allows two regimes to be distinguished: a) below $z$ = 0.35-0.40 the surface of the central feature increases with $z$; b) above $z$ = 0.4 a plateau has been reached.
In the low $z$ bin, there is a competing effect between resolution and S/N in cluster cores and within the  overlapping range the basic and redshifted samples are indistinguishable. In region b) scale 3 includes  an increasing fraction of the cluster outskirts, but the number of photons is decreasing with $z$ yielding the plateau for the size of scale 3.
In Fig. 12 \& 13  we have further plotted   the morphology parameter  as a function of $z$ for both samples. The two distributions are very similar and no correlation is found between $z$ and morphology . 
Thus, the geometrical characteristics of the central feature does not seem to provide a handy test. 
We have however investigated any dependence of the central morphology on the  mass flow rate: no correlation is obvious  (Fig 14). This averaged result however, does not  exclude the presence of peculiar features in the core of some clusters as shown in Sec. 4.2

 \begin{figure}
 \psfig{file=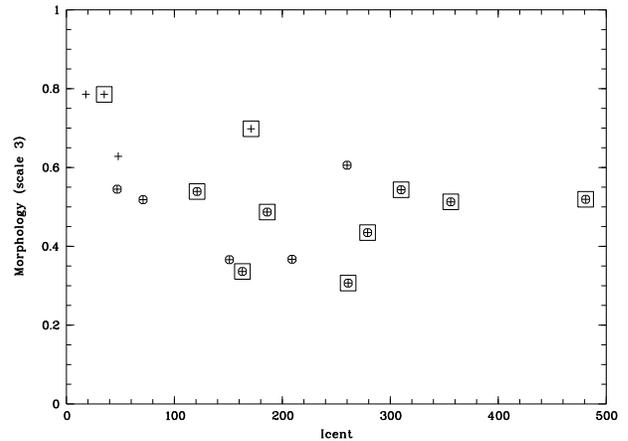,width=9cm,angle=-90}
 \caption[] { The central morphology as a function of the central photon numbers. Points having a central structure with more than 10 significant pixels are encircled. Cooling flow clusters are indicated with a square.
 The two leftmost (and uppermost) points correspond to clusters for which $I_{cent} \leq   47 $ i.e. below the detectability threshold of scale 3. }
\end{figure}

 \begin{figure}
\psfig{file=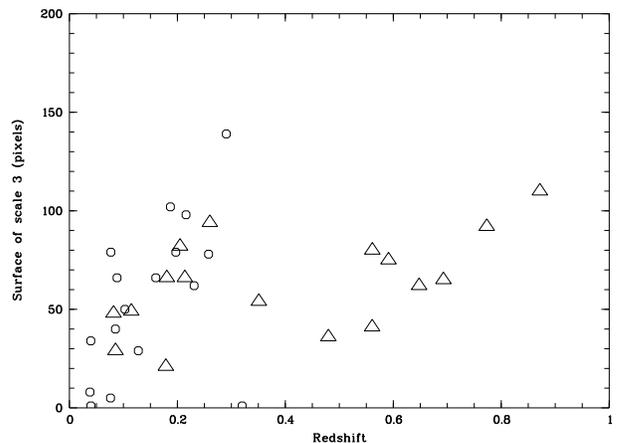,width=9cm,angle=-90}
 \caption[] {Surface of the central structure at scale 3 as a function of z. Circles: basic sample; triangles: redshifted sample. }
\end{figure}

 \begin{figure}
\psfig{file=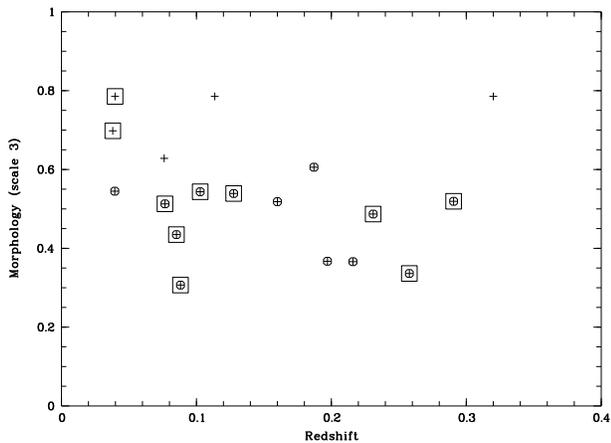,width=9cm,angle=-90}
 \caption[] {The morphology parameter as a function of redshift for the scale 3 (crosses). Same symbols as Fig. 10.}
\end{figure}

 \begin{figure}
 \psfig{file=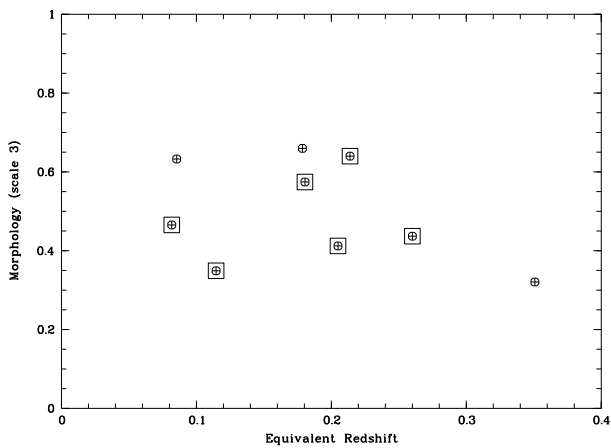,width=9cm,angle=-90}
 \caption[] {Same as Fig. 12, but for the redshifted sample. Both distributions are basically indistinguishable.}
\end{figure}

 \begin{figure}
\psfig{file=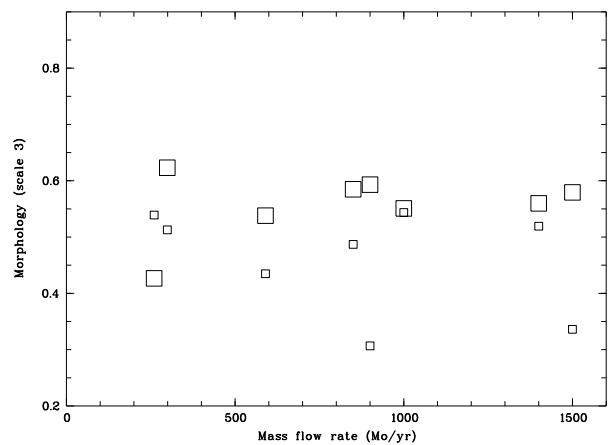,width=9cm,angle=-90}
 \caption[] { Cooling mass flow rate as a function of central morphology. A circle would have a morphology parameter of 1. Small squares correspond to scale 3 and large ones to scale 4. Only objects having a central feature extending at least over 10 pixels are displayed. }
\end{figure}

We have investigated possible variations of the  isophotes PAs as a function of radial distance (so-called  ``twisting isophotes'' ) in our sample: Fig. 15 \& 16. Ellipses were fitted  to the wavelet filtered images (see Fig. 5)   according the method of Bender \& Moellenhoff (1987) which allows for the removal of multiple minor perturbations by comparing the values of pixels at opposite positions angles (the fit was performed on the 3 and 3.7 $\sigma$ restored images, yielding extremely comparable results as to the content of Fig. 15 \& 16).
On the background subtracted images, position angles were measured for two different radial distances -- PA1, PA2 -- corresponding to a fall off of   the central luminosity of $I_{1} = 10^{-0.2}I_{o}$ (usually a few pixels off the center)  and $I_{2} = 10^{-1}I_{o}$ respectively (when taking $I_{1} = 10^{-0.45}I_{o}$, the following correlations  almost disappear). In this way,  we are addressing similar regions from cluster to cluster over the whole redshift range.
Clusters included in the sample were those for which fitting occured to be possible at these two extreme radial values (flagged in Table 2).

  \begin{figure}
\psfig{file=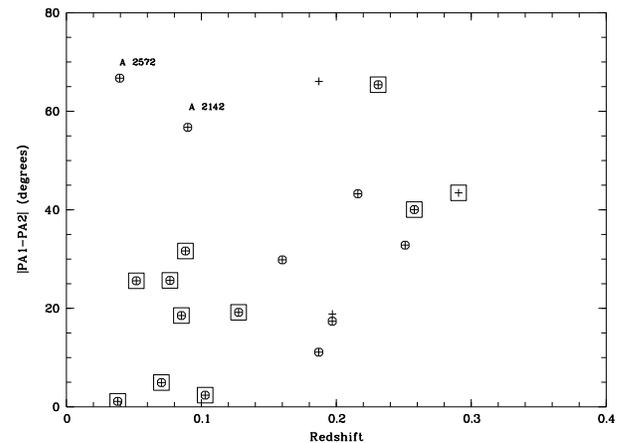,width=9cm,angle=-90}
 \caption[] { Variation of the position angle for two given isophotal levels (external-internal) as a function of redshift for the basic sample. Cooling flow clusters are indicated with a square. The two labeled clusters are  perturbated by the presence of a close neighbor, which explains the observed large PA variations. Encircled points correspond to objects having an excentricity   $\ge 0.5$ at $I_{1}$. }
\end{figure}

\begin{figure}
 \psfig{file=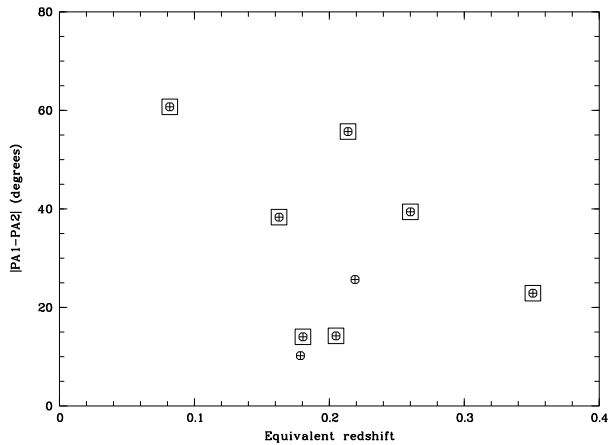,width=9cm,angle=-90}
 \caption[] {Same as Fig. 15 but for the redshifted sample }
\end{figure}


Except for two objects: A2572, A2142 (which have close neighbors) there is a clear correlation between the twisting angle and $z$. A similar trend was also observed using PA of the wavelet scales 3 \& 6.  This effect is not present in the redshifted sample, and could thus be considered as an intrinsic trend.
No obvious correlation with morphology was found between the presence of a cooling flow or the profile steepness (ratio of the ellipse major axes at $I_{1}$ and $I_{2}$ and the twisting angle $\| PA_{1}-PA_{2} \|$).

 In contrary  to previous claims, very few isolated pointlike   features and even fewer ``filaments" are detected in the cluster cores
 at the 3$\sigma$ level, in addition to central feature. Out of these, only a few of them remain at    3.7$\sigma$;   the missing features could be considered as spurious detections, inherent in the method (cf paper I). This can be easily checked by comparing Fig. 6 (wavelet images at the $3\sigma$ level) and Fig. 5 (reconstructed filtered images at the $3.7\sigma$ level). These peculiar small scale features are discussed next section. \\

\subsection{Notes on structures found in individual clusters}
We present here a detailed review of the cluster core morphologies.
The presence of  pointlike sources is discussed in connection with possible optical counterparts (digitized sky surveys; Cosmos http:\\xweb.nrl.navy.mil/www\_rsearch/RS\_form.html; APM http:\\www.ast.cam.ac.uk/~apmcat) only for  those present in the 200 $\times$ 200 kpc$^{2}$ central area and having a 3.7 $\sigma$ significance (see Fig. 5 \& 6). In the optical/X correlation,  the X-ray positions are those directly derived from the ROSAT attitude solution. The core of several clusters - beyond $z\sim 0.1$ - appear to be significantly twisted with regard to isophotes at larger radial distance which show, in general, an orientation comparable to that of the cD.\\

{\bf A496} The HRI image was first studied by Prestwich et al (1995): after ellipse fitting, two residuals were found. In agreement with their results, in our analysis only residual X1 is ascribed a significance of at least 3$\sigma$. As 171 photons were found in the central box but no significant signal at scale 3 for the cluster center (maximum at scale 5), it seems very likely that the central region of this cluster is here totally resolved 
The small feature 15" off the center (X1) is contained within the cD envelope. \\
{\bf A2572 S} This cluster is truly A2572 at z = 0.0395 (Ebeling et al 1995). Our analysis shows strong evidence for a pointlike contribution at the cluster center (the wavelet coefficient is maximum - and by far -  at scale 3).\\
 {\bf A3571} No significant structure has been found at the cluster center at scales 3 and 4, but the signal in the central box is rather low: 35 photons. For the central region, the wavelet coefficient is maximal at scale 6. According to the Cosmos data base, the central feature and that some 30'' eastward are both contained within the cD envelope. There is a pointlike object (b=15) some 4" from the source located at x=-100, y=250. The overall orientation of the elongated X-ray core is very similar to that of the central galaxy ($\leq$ 10 deg) .\\
{\bf A85} Like in A496,  several cooling ``filaments" were found by   Prestwich et al (1995), in the HRI image of A85 but all strictly below 3$\sigma$. Out of these, only residual X1 was detected by our method as a slight  SE elongation of the plane 4 contour, thus clearly resolved (consequently, this is not a filament). In addition, north to the cluster center, we find two point-like sources having a significance $\leq$ 3.7 $\sigma$ (not in the field displayed by Prestwich et al) : the source at [-450,-350] is only at 1.08' from the cD and no galaxy brighter than b=20 is found within a radius of 1.0'. There is a galaxy with b=17.0 at 6" from the source located at [-250,-400]. Like for A0496, the scale 4 contour is within the cD envelope (Cosmos). \\
{\bf A644} The central emission is faint; it is, thus, difficult to determine the X-ray maximum but there is a clear indication of twisting isophotes. The geometrical center of the central elongated feature correspond to the cluster dominant galaxy (Cosmos).
A644 is a lens cluster (Maurogordato et al 1996). \\
{\bf A1423} There is faint central signal at scale 3 ($I_{cent}=48$), higher, however than at scale 4: there is thus a clear indication for a pointlike contribution. The west elongation of scale 4 is contained within the cD enveloppe (APM). The cD has an orientation comparable to that of  scale 4. \\
{\bf A2029} The HRI image of this cluster has been subject of controversy: filaments were found by Sarazin et al  (1992 a,b).
Subsequent analysis by Thompson \& Prestwich (1994) found it to be ``consistent" with Poisson noise, as well as White et al (1994) who did not find substructures in the form of significant departure from smooth elliptical isophotes. Our analysis, shows indeed a very regular cluster and some indication of a non resolved structure at the very center.\\
{\bf A2597} Sarazin et al (1992a) mentioned inhomogeneous emission in the HRI image; no structure was found by Thompson \& Prestwich (1994). Our analysis suggests some possible unresolved contribution at the cluster center. The elongation of the X-ray core has been discussed in detail by Sarazin et al (1995) in connection with radio observations.\\
{\bf A478} This image has been previously studied by White et al (1994).
Our analysis shows a very elongated (not resolved) core, with possible 2 maxima; its PA is significantly twisted with respect to the large scale cluster orientation which is comparable to that of the cD ($\sim 35$ deg. APM) . The elongated central structure is confined within the extent of the cD envelope.  \\
{\bf A2142} In the late state of a merger (Henry \& Briel 1996). It was already flagged by Buote \& Tsai (1996) as typical ``off-center" cluster. Our analysis tends to confirm their conclusion reached from the PSPC data analysis, as several small scale structures are conspicuous.\\
{\bf PKS 0745-191} This cluster presents a case of strong lensing and has a massive cooling flow (Allen et al 1996b). The north-west  small scale feature has a significance between 3.0 and 3.7 $\sigma$. Its presence had already been noticed by Allen et al. Here, it is clearly separated from the central structure with an offset of 11" and 8" in RA and Dec respectively, which places it only some 5" from the giant arc. There is no optical identification obvious exactly at this place, but a pointlike object about 5" south of it is visible in the CCD image of Allen et al. This structure is also contained within the cD extended emission and thus could well be associated to the CF. The configuration involving the cD,  the arc, the pointlike feature (all aligned) is very comparable to that found in another massive CF cluster: A2390 (Pierre et al 1996). \\
{\bf A1664} The cluster presents a strong central elongation at scale 4, clearly off-centered, with respect the scale 3 (centered on the cluster dominant galaxy and with comparable orientation). This would indicate that the central region may be composed of two rather large clumps. But on larger scales the centroid shift is somewhat smoothed out (Fig. 5) and found to be ``normal" for this redshift (Fig. 8).  \\
{\bf A2580} A number of $I_{cent}$ = 63 photons was found at the center, but no structure at scale 3; thus, at this intensity level, we can exclude a significant contribution from a central unresolved feature. \\
{\bf RXJ2318.3+1843 (A2572 N)} This cluster located only 3.2' from A2572  is actually a background object at $z \sim 0.16$ (Ebeling et al 1995), but having a higher flux. \\ 
{\bf A383}  An apparently well relaxed object. The X-ray emission is centered
on the cluster dominant galaxy and the bright source some 1' North-East from the center may be associated to a pointlike object (b=20.3 at $\sim$ 8", Cosmos) in the optical. \\
{\bf A773} Two separate maxima are found at the cluster center separated by some 35''. According the APM data base, the galaxy density is high at this place, but none of the X-ray peaks can be unambiguously identified with a  given galaxy. \\
{\bf A115} The curved shape of the central feature  could be due to the presence of two distinct pointlike sources; it seems hard to attribute the presence of the ``hole" to photons statistics, since it has a size comparable to that of the PSF; it is also seen in the two separate pointings which constitute the observation. According to the Cosmos data base the central feature has an orientation comparable to that of the cD, and totally contained within its envelope. \\  
{\bf MS 0735.6+7421} Evidences were found for the presence of a cooling flow in this distant object by Donahue \& Stocke (1995) using optical, PSPC and HRI data, but they did not provide an estimate for the mass flow rate. They performed a somewhat detailed morphological analysis, which did not reveal significant structure on scale 20"-230". Our multi-resolution analysis is consistent with this but shows, in addition, a very elongated core with strong evidences for the presence of two maxima (confined within the cD enveloppe). 
The core of the cluster, is clearly not aligned with scale 5 and at $I_{2}$ we find a PA of -30 deg which is close to that of the cD (-21 deg on plate 680, according to APM) \\
{\bf A2390} The properties of this massive cooling flow - and lensing - cluster have been studied in detail
by Pierre et al (1996). A slight rectification is needed: the significance of residual possibly associated to galaxy \# 314 is between 2.3-3 $\sigma$ (and not 3$\sigma$ at least, as stated in the 1996 paper). The central elongated feature has also a size comparable to that of the cD but their position angle are very different:
the PA of the cD is -49 deg (Pierre et al 1996) and that of the core $\sim $ 90 deg. On larger scales we found PA($I_{2}$) = -56 deg which is comparable to that of the cD .\\
{\bf A2645} This cluster has a low central intensity ($I_{cent}=20$). \\
{\bf E1455+223}  This cluster  was observed by ASCA and ROSAT (Allen et al 1996a) and was shown to host a massive cooling flow. Our analysis reveals clearly the presence of two non-resolved point sources at the cluster center (3.7$\sigma$ level at least); since the HRI image we used is a merge of three separate pointings (92Jan, 93Jan, 94Jul) we checked that the double source is not an artifact due to small pointing offsets: this is not the case and, actually, the presence of the double source is conspicuous in at least two individual pointings. According to the APM data base, the two maxima are  located within the cD envelope and, thus, very probably related to the cooling flow. Again here - as can be seen on Fig. 5 - there is a clear twist between the inner core and scales 4-5. But like for MS 735 and A 2390 we find a good agreement between alignment of the cD (PA is 36 deg. APM) and  outer scale (PA  at $I_2$  is 39 deg.)\\
{\bf Zw 3146 (ZwCL 1021.0+0426)} This cluster has an extremely peaked emission: despite its redshift, it presents the highest central intensity (cf Table 1, col. Icent). This explains the strength of the observed signal on scale 3. The emission
appears however to be resolved since the wavelet coefficient is maximal at scale 4. The wavelet analysis suggests the presence of two secondary maxima in the immediate vicinity (S and NE from the center). This image is made of two separate pointings, each showing a clear NS elongated core. \\
{\bf S506 (CL 00500-24)} Well known lens cluster (Fort \& Mellier, 1994 and reference therein). The peak of scale 3 corresponds to the cluster dominant galaxies (double) within 3" (Cosmos). The X-ray emission is very faint and results on structures should be considered with caution. The gaussian filtered image is discussed by Schindler \& Wambganss (1997). We find confirmation with our method of their secondary north-east maximum at a signifcance between 3-3.7 $\sigma$  (the feature is outside the 200 $\times$ 200 Mpc field of Fig. 6.)\\

\section{Discussion}

Cluster cores are thought to be the place where virialisation  first occurs and thus in this respect, should present an overall smooth distribution of the X-ray emitting gas. 
However, in cooling flows - and most probably in the whole ICM - the presence of small scale inhomogeneities is expected as a result of the  development  of thermal instability (e.g. Nulsen 1986). (Peculiar emission from individual galaxies may be also observed, although at the redshifts of interest in the present paper ($ \geq 0.04$)  - and  S/N -
such a positive detection would be most certainly due to an AGN.)
It is thus of prime interest to statistically investigate at the finest possible resolution, the very center of a representative sample of clusters, in terms
of luminosity, redshift and  strength of the cooling flow. 

Using ROSAT HRI images, we have attempted to characterize the shape of cluster
cores, their relation to the rest of the cluster and to search for small scale
structures. The X-ray images were analyzed by a wavelet method combined to a rigorous treatment of the local photon number fluctuations. We have built from the original images a ``redshifted sample", to check evolutionary trends. 

We can summarize our findings in the following way:\\
- In terms of shape (see Sect. 4.1) of the smallest central scale, we find no significant difference between, CF and non CF clusters, low and high z clusters.\\
- In terms of isophote orientation and centroid shift, two distinct regions  appear and seem to co-exist: the central inner 50-100 kpc and the rest of the cluster. We find a clear trend for less relaxation with increasing $z$.\\
- In general, very few isolated clumps are detected above $3.7 \sigma$  in the cluster central region out to a radius of $\sim 200 $ kpc. Peculiar central features have been found in a few high $z$ clusters.\\

\subsection{ Relation with cooling flows}

In order to understand the morphology/evolution results obtained above, it is  necessary to  underline the properties of our sample in terms of CF. First, as shown by Fig. 4    there is a clear  trend for massive CF clusters to be located at high $z$. Whether this is universal is delicate to assess sice  moderate CF clusters are less obvious to detect at high redshift. Alternatively,  the status of our sample may simply reflect observing selection biases, but it is significant that no low $z$, very massive CF cluster is known. The fact that strong CF clusters may exist at early epochs was already suggested  by Henriksen (1993) allowing for CF to form in rich clusters after the cluster merge to form a single potential well but before galaxies have reached equilibrium.\\
In our analysis, CF clusters tend to show some structure within a scale comparable to that of the cD envelope, which means at scale 4 below $z \sim 0.09$ and scale 3 above. This is apparently true for the whole redshift range of interest but also for the  redshifted sample. Indeed, the cores  of the $redshifted$ clusters A496 \& A85 closely resemble  those of A2029, A2597, A478 \& PKS 745 although the latter ones have significantly higher mass flow rates. On a scale smaller than the cD (e.g. scale 3 for the low $z$ objects) only one significant structure is found ( 3.7 $\sigma $) - in A3571; unfortunately these clusters are also the weakest in terms of CF. Also, none of the 4 nearest CF clusters   show unresolved feature at the very center, whereas at least two of them (A496, A85) have enough S/N to enable such a detection. Moreover,   A115 presents a peculiar central structure but is not known as CF cluster. So, in summary, our analysis  -  despite having revealed so far unsuspected central structures in at least two objects (MS 0735.6+7421, E1455+233) - shows that with the present angular resolution and sensitivity,  CF and non CF cluster cores are indistinguishable in terms of global shape. \\

\subsection{Relation to external parts}
There are a number of important points to consider on larger scales, especially in the comparison of the cluster cores and outer regions.\\
Cases are found where the orientation of the cD is comparable to that of  
scale 3 (e.g. A1664) and others where the two directions are significantly different,  although on larger scales both PAs agree well (e.g. the strong CF cluster A2390).  Cases of alignment have been previously pointed out, for instance by Allen et al (1995) investigating clusters in the $0.1\leq z \leq 0.15 $ range. They found an agreement within 5 deg. for CF clusters within $r\leq 0.5$  Mpc, and poorer for a non CF cluster. Here, for  A2390 ($z= 0.231$)   we found a large misalignment ($\sim$ 45 deg) between scale 3 and the cD whereas for scales greater, the agreement between the cD and the X-ray gas is good. This could indicate  a non totally relaxed state as suggested by the galaxy distribution and the gas clumpiness on larger scales (cf Pierre et al 1996).  Alternatively,  it may mean that local processes connecting the hot gas and the cD itself are dominant.  The cases of MS 0735.6+7421 and E1455+233 appear  similar. \\ 
 The phenomenon of twisting isophotes was already well known for light profiles of elliptical galaxies. This seems to be also a rather common feature of X-ray gas profiles in clusters (e.g. Mohr et al 1993) and is usually interpreted in terms of unresolved substructures within the ICM; alternatively, de Zeeuw and Carollo (1996) have shown that (for galaxies) simple triaxial mass models with a central density cusp can indeed show ellipticity variations and isophote twist.
We have found (Fig. 15) a significant increase with redshift of the mean PA variation between two given isophotal levels.  The dependence of this effect on the CF strength is not straightforward to assess since most
of the clusters for which the measurement was possible are CF objects. However, the effect is not present in the redshifted sample and, as no correlation between the presence of a CF (or the profile steepness) and the twisting angle was found, it is  likely of cosmological origin. This could well indicate that the PA variation as defined here is a good indicator of the cluster relaxation state.  

Using a power ratio technique applied on  classical CDM simulation clusters, Tsai and Buote (1996) have shown that there is a continuous competition between relaxation and formation rate along the whole cluster history. They do not observe a significant change up to $z \sim 0.6$, where both effects come into balance; after this time, the formation rate immediately levels off.  Their analysis was based on simulated PSPC images, integrating counts out to  $r=1$ Mpc, i.e. sensitive to rather large scale features and possibly peripherical. Here, HRI data combined with our wavelet analysis exclude investigating external regions but allow a much finer study of the cluster inner core ($r = 50-100$ kpc) and of its relations to the outskirts. Indeed, we found in terms    of ``centroid shift" and ``isophote twisting" a significant increase from present to $z \sim 0.4$. Since the effect is not present in the redshifted sample we can infer that there is an intrinsic difference between the two samples, which means actually between clusters below and above $z \sim 0.1$. 
There is a large scatter in the correlation, probably partly internal and partly due to a combination of projection/orientation effects.\\

Our present results suggest that there is good observational evidence for the very  inner cluster regions  to show on scales of 50-100 kpc - in the X-ray band -  peculiar properties with respect to more external parts. Indeed, this scale is comparable to that of the cD, which is by no means a ``normal" galaxy. These giant elliptical galaxies are supposed to occupy the center of the cluster potential, the place which first should become virialized. X-ray informations confirm this since -- except for a very limited number of cases (e.g. the group AWM7, Neumann \& B\"ohringer 1995) -- the maximum of the emission appears to be always located at the cD center (when the X-ray astrometric solution is confirmed by additional pointlike sources). This is also the place where the ``cooling flow" gas accumulates (if we assume the conventional CF picture) and must at some stage be processed into stars within the cD. It is thus a favored place for the interplay between the ISM and the ICM, although observational constraints as to the fate of the cooling gas are still rather loose. When considering the more elaborated view of a multi-phase medium in which the flow is actually substantially affected   by the deposit of condensations (initiated by thermal fluctuations) we probably better understand why so few fluctuations have been detected. Blobs of cold and dense material do exist but their individual emissivity is  actually averaged in projection and only shows up  in the spectral analysis (cf Allen et al 1996a).  The inner 50-100 kpc scale is also comparable to the lensing core radius, and it has been shown that models reproduce strong lensing configuration at best
when the ellipticity and orientation of the dark matter distribution  is taken to follow that of the cD  (Fort \& Mellier 1994).  However, central mass estimates from lensing analysis tend to show that mass profiles are much steeper than that derived from X-rays alone (when temperature information is available) which is itself steeper than the gas mass distribution. Here, the cD/X-ray alignment discrepancy found here in several cluster cores (some of them being strong lenses) suggests also that the gas may not follow the dark matter distribution on the cD scale.

In other words, the present study, down to the  limiting instrumental resolution, enables us to isolate - in terms of dynamical and physical state 
- central regions down to a scale comparable to that of the cluster dominant galaxy. However it was not possible to infer firm connexions between central morpholgies and  cooling flow rates or redshift.
Our results allow us to witness for the first time at the cluster center, the competition with the relaxation processes which should here be well advanced and  local phenomena due to the presence of the cD galaxy.  Forthcoming AXAF and XMM observations at much higher sensitivity, over a wider spectral range and with a better spatial resolution may considerably improve our understanding  of the 
multi-phase plasma and of its inter-connections  with the interstellar medium.



\begin{thebibliography}{}
\bibitem{} Allen S.W., Fabian A.C., Edge A.C., Bautz M.W., Furuzawa A., Tawara Y., 
1996a MNRAS, 283, 263
\bibitem{} Allen S.W., Fabian A.C., Kneib J.P., 1996b  MNRAS, 279, 615
\bibitem{} Allen S.W., Fabian A.C., Edge A., B\"ohringer H., White D. A., 1995  MNRAS, 275, 741
\bibitem{} Allen S.W, Fabian A.C., Johnstone R.M., White D.A., Daines S.J., Edge A.C.,Stewart G.C., 1993 MNRAS, 262, 901 
\bibitem{} Bender R, Moellenhof C, 1987 A\& A, 177, 71
\bibitem{} Bird C. M., 1994  AJ, 107, 1637 
\bibitem{} Briel U.G., Henry J. P., Schwarz R.A., B\"ohringer H., Edge A.C., Hartner G.D., Schindler S., Tr\"umper J., Voges W., 1991 A\&A, 264, L10
\bibitem{} Buote D., Tsai J, 1995 ApJ, 452, 522
\bibitem{} Buote D., Tsai J, 1996 ApJ, 458, 27
\bibitem{} Crone M. M., Evrard A. E., Richstone D. O., 1996 ApJ, 467, 489
\bibitem{} David L.P., Harnden F.R., Kearns K.E., Zombeck M.V., Harris D.E., Prestwich A., Primini F.A., Silverman J.D., Snowden S. L. 1997, {\em The ROSAT HRI Calibration report}, http:\\hea-www/harvard.edu/rosat/rsdc\_www/HRI\_CAL\_REPORT/hri.html.
\bibitem{} De Zeeuw P. T., Carollo C. M., 1996 MNRAS, 281, 1333
\bibitem{} Dickey J. M., Lockman F. J. 1990 AN. Rev. Astron. Astrophys.m 28, 215
\bibitem{} Donahue M. \& Stocke J.T., 1995 ApJ, 449, 554
\bibitem{} Dutta S.N., 1995 MNRAS 276, 1109
\bibitem{} Ebeling H., Voges W., B\"ohringer H., Edge A.C., Huchra J. P., Briel U. G., 1996 MNRAS, 281, 799
\bibitem{} Ebeling H., Mendes De Olivera C., White D. A., 1995 MNRAS, 277, 1006
\bibitem{} Edge A.C, Stewart G.C., Fabian A.C., 1992 MNRAS, 258, 177
\bibitem{} Fort B., \& Mellier Y., 1994 A\&A Rv, 5, 239
\bibitem{} Grebenev S.A., Forman W., Jones C., Murray S., 1995 ApJ  445, 607 
\bibitem{} Henriksen M. J., 1993 ApJ, 407 L13-L15
\bibitem{} Henry J. P., Briel U. G., 1996 ApJ, 137, 144
\bibitem{} Lemonon L., Pierre M., Hunstead R., Reid A., Mellier Y., B\"ohringer H., 1997 A\&A in press
\bibitem{} Maurogordato S., Le F\`evre O., Proust D., Vanderriest C., Cappi A., 1996 CFHT Bulletin 34, 5
\bibitem{} Mohr J.J., Fabricant D. G., Geller M. J., 1993 ApJ, 413, 492
\bibitem{} Neumann D. M., B\"ohringer H., 1995 A\&A, 301, 865
\bibitem{} Neumann D. M., B\"ohringer H., 1997 MNRAS in press, Astro-ph/9607063
\bibitem{} Nulsen P. E. J., 1986 MNRAS, 221, 377
\bibitem{} Pierre M., Le Borgne J.F., Soucail G., Kneib J.P, 1996 A\&A 311, 413
\bibitem{} Prestwich A.H., Guimond S.J., Luginbul C.B., Joy M., 1995 ApJ, 438, L71
\bibitem{} Richstone D., Loeb A., Turner E.L., 1992 ApJ 393, 477
\bibitem{} Sarazin C.L., O'Connel R.W., McNamara R.R, 1992a BAAS, 181, 1501
\bibitem{} Sarazin C.L., O'Connel R.W., McNamara R.R, 1992b ApJ, 389, L59
\bibitem{} Sarazin C.L., Burns J.O., Roettiger K., McNamara B.R., 1995 ApJ, 447, 559
\bibitem{} Schindler S., 1996, A\&A, 305, 756\\
\bibitem{} Schindler S. \& Wambganss J., 1997 A\&A, 322, 66
\bibitem{} Slezak E., Durret F., Gerbal D., 1994 AJ, 108, 1996
\bibitem{} Tompson H., Prestwich A, 1994 BAAS, 185, 7403
\bibitem{} Tsai J., Buote D., 1996 MNRAS, 282, 77 
\bibitem{} White D.A., Fabian A.C., Allen S.W, Edge A.C., Crawford C.S., Johnstone R.M., Stewart G.C., Voges W., 1994 MNRAS, 269, 589.
\end{thebibliography}
\end{document}